\documentclass[fleqn,usenatbib]{mnras}

\usepackage[T1]{fontenc}
\usepackage{ae,aecompl}
\usepackage[draft]{hyperref}

\usepackage{graphicx}	
\usepackage{amsmath}	
\usepackage{amssymb}	

\def\be{\begin{equation}}
\def\ee{\end{equation}}

\DeclareMathOperator{\erf}{erf}

\title[Analytical model for cluster radio relics]{Analytical model for cluster radio relics}
\author[M. Br\"uggen, F. Vazza]{M. Br\"uggen$^{1}$\thanks{E-mail: mbrueggen@hs.uni-hamburg.de},
F. Vazza,$^{2,1,3}$
\\
$^{1}$University of Hamburg, Hamburger Sternwarte, Gojenbergsweg 112, 21029 Hamburg, Germany \\
$^{2}$Dipartimento di Fisica e Astronomia, Universit\'{a} di Bologna, Via Gobetti 93/2, 40121, Bologna, Italy\\
$^{3}$Istituto di Radio Astronomia, INAF, Via Gobetti 101, 40121 Bologna, Italy
}

\date{Accepted XXX. Received YYY; in original form ZZZ}

\pubyear{2015}

\begin{document}
\maketitle

\begin{abstract}
Radio relics are vast synchrotron sources that sit on the outskirts of merging galaxy clusters.
In this work we model their formation using a Press-Schechter formalism to simulate merger rates, analytical models for the intracluster medium and the shock dynamics, as well as a simple model for the cosmic-ray electrons at the merger shocks.
We show that the statistical properties of the population of radio relics are strongly dependent on key physical parameters, such as the acceleration efficiency, the magnetic field strength at the relic, the geometry of the relic and the duration of the electron acceleration at merger shocks.
It turns out that the flux distribution as well as the power-mass relation can constrain key parameters of the intracluster medium. With the advent of new large-area radio surveys, statistical analyses of radio relics will complement what we have learned from observations of individual objects.    
\end{abstract}




\section{Introduction}

Radio relics are diffuse radio sources in galaxy clusters. They show radio synchrotron emission with steep spectra and they appear to trace shock waves that form in the course of cluster mergers (see \citealt{2019SSRv..215...16V} for a recent review).
Here we focus on radio relics that are caused by shock waves, also called cluster radio shocks. Other types of relics, such as the so-called phoenix sources are even less well understood and not the subject of this paper.
Merger shocks have been detected in X-ray observations, for example in the Bullet Cluster (1E 0657-56) \citep{2002ApJ...567L..27M} but also in other cases \citep[e.g.]{2005apj...627..733m, 2014mnras.440.3416o, 2016MNRAS.458..681D, 2015a&a...582a..87a}. These shocks are typically located at distances of $\sim 1$ Mpc from the cluster centres, and they have low sonic Mach numbers, in the range $M\sim 1.2 - 4$.
Note that the Mach numbers inferred from X-ray observation are often lower than those inferred from the radio spectral index assuming diffusive shock acceleration, for example, in the Toothbrush relic \citep{2015pasj...67..113i,2019arXiv191108904R} and in the radio relic in A2256. 
Even though such giant radio relics are not very common, the fraction of X-ray luminous clusters hosting relics is estimated to be around 10\% \citep[][]{2019SSRv..215...16V}.
Models where electrons are accelerated at merger shocks via diffusive shock acceleration have been able to reproduce some of observed properties of relics such as their overall flux, their morphologies, and the radial profiles of their spectral indices \citep{2012apj...756...97k, 2017apj...840...42k, 2018apj...857...26h}.

Particularly interesting, are so-called double radio relics, where two convex radio sources are found on opposite sites of the cluster centre \citep[e.g.][]{2014mnras.444.3130d}. X-ray observations show that these relics are oriented perpendicular to the major axis of the distribution of the intracluster medium, and the presumed merger axis. These shocks are sometimes called axial shocks \citep[][]{2018apj...857...26h}.  Interestingly, double radio relics seem to be observed close to edge-on and are believed to trace shocks that were presumably triggered at core passage of the progenitor clusters \citep[][]{2019SSRv..215...16V}.

\citet{2015apj...805..143d} and \citet{2017arxiv171101347g} conducted an exhaustive analysis of a sample of 29 relics and found that the merger axes of relics lie near the plane of the sky. This suggests that relics are affected by limb brightening of the emission in the thin shock waves and may lead to many relics with less favorable orientations being missed in current samples.

The radio power of radio relics has been found to correlate with the X-ray luminosity of the host cluster \citep[][]{2012a&arv..20...54f}. \citet{2014mnras.444.3130d} found a correlation between cluster mass and the specific radio power, which goes as $P_{1400} \propto M^{2.8}$. This correlation is biased because all the low-mass clusters are very close and the absence of low-power radio relics in massive clusters is affected by Malmquist bias. As a result, especially at the low-mass end, the phenomenon of radio relics is poorly understood \citep[e.g.]{2017a&a...597a..15d}.\\

Moreover, there is a correlation between the largest linear size and the distance from the cluster centre \citep[e.g.]{2009a&a...503..707b}, even though all these correlations are based on small samples without clear selection criteria. There have been some theoretical efforts to explain these correlations, finding some agreement with the properties of relics found in the NVSS survey \citep{2012mnras.420.2006n, 2012mnras.423.2325a, 2017mnras.470..240n}. An updated list of radio relics can be queried under \href{http://galaxyclusters.com}{http://galaxyclusters.com}.\\

More recently, \cite{2019arXiv190700966B} have studied 10 radio relics that have shocks confirmed in the X-rays. Using the Mach numbers derived from the X-ray data, they calculated the required electron acceleration efficiency to reproduce the observed radio luminosity. They conclude that other mechanisms, such as the re-acceleration of fossil electrons, are needed to explain the origin of radio relics. This confirmed prior work done by \cite{2014mnras.437.2291v, 2013mnras.435.1061p, 2005apj...627..733m}.
Nearly all of the merger shocks are found in systems where the centres of the clusters have already passed each other. However, recently \cite{2019NatAs.tmp..375G} have discovered a shock in a pre-merger phase in the cluster pair 1E2216 and 1E2215.\\

Radio relics are particularly interesting as they also represent a laboratory for studying particle acceleration in low-Mach number shocks. The low-density, high plasma-$\beta$ (i.e. $\beta \sim 10^2$) environment of the intracluster medium (ICM) cannot be modelled in laboratories on Earth and hence there are only numerical simulations to compare to. 

Structure formation shocks have been investigated in a number of semi-analytical \citep{2003ApJ...594..709B, 2003apj...585..128k} and numerical papers \citep{2000apj...542..608m, 2001apj...562..233m, 2003apj...593..599r, 2005ApJ...620...44K, 2008mnras.385.1211p, 2008A&A...481...33J, 2008MNRAS.391.1511H, 2009MNRAS.393.1073B, 2011apj...735...96s, 2013apj...765...21s, 2017mnras.464.4448w}. 

In previous work, \cite{2007mnras.375...77h} have produced a semi-analytical scheme for modelling individual relics if the shock area and the magnetic field are known. \cite{2008MNRAS.391.1511H} studied diffuse radio emission from galaxy clusters using the cosmological Smoothed-Particle Hydrodynamics (SPH) simulation "MareNostrum". They found that the maximum luminosity of diffuse radio sources in galaxy clusters scales with their X-ray temperature. Moreover, they concluded that accretion shocks around galaxy clusters give rise to only very weak radio emission. Their model shows that a moderate shock acceleration efficiency of, of 0.5\%, and magnetic fields of 0.07-0.8 $\mu$G, suffice to reproduce the number density and luminosity of radio relics.  Using the same simulation, \cite{2012mnras.423.2325a} found that 41\% of clusters with masses $M >10^{14}M_\odot$ host radio relics. Using adaptive-mesh refinement simulations, \cite{2011apj...735...96s} investigated scaling relations between radio luminosity and cluster properties such as mass and X-ray luminosity. 

Using a cosmological simulation, \cite{2012mnras.420.2006n} have related the halo mass function to the radio-relic number counts. They estimate that the Westerbork Observations of the Deep APERTIF Northern-Sky (WODAN) survey proposed for WSRT could discover 900 relics and that the LOFAR-Two Metre Sky Survey (LOTSS) may discover about 2500 relics. They also found scalings of the radio power of relics with the X-ray luminosity of the host clusters.\\

Using zoomed galaxy cluster simulations, \cite{2017mnras.470..240n} have shown that simulated relics can partly match observations of relics from the NVSS survey. However, they find that number counts in the simulated relic sample increase more steeply towards lower flux densities, which indicates that the NVSS is not complete at this end. They also find disagreements with observed properties, in particular a mismatch for the sizes and projected distances of relics in distant clusters at $z>0.3$. Several inferred properties of radio relics remain mysterious, in particular the seemingly high acceleration efficiency at low Mach-number shocks as well as the apparently high magnetic fields inside the relics.

\citet{2003ApJ...583..695G} have used a similar approach that is used in this paper by constructing merger trees that are then used to infer Mach number distributions. They find that only minor mergers produce the high Mach numbers required for efficient particle acceleration. This is not confirmed by observations that show that major mergers between very massive clusters mainly host relics and halos \citep[see e.g.][]{2001apj...553l..15b}. 

During the mergers, also protons get accelerated that stay confined in the cluster volume for cosmological time scales. These protons collide inelastically with thermal protons from the ICM, producing neutral pions that decay emitting gamma rays. Thus, they can contribute high-energy gamma rays. \citet{2003ApJ...583..695G} have shown that inverse Compton scattering of electrons in galaxy clusters cannot contribute more than about 10\% of the observed diffuse gamma-ray background. This number has been confirmed by numerical simulations, e.g. from \citet{2001apj...562..233m}. 
We will only focus on the acceleration of electrons and sidestep the role of protons for gamma-ray observations of galaxy clusters.

In this paper, we devise an analytical model for the luminosity function of radio relics and show how physical parameters of this model affect the resulting statistics of radio relics. By producing flux and power distribution functions and correlations between host mass and radio power, our results can be compared to data on radio relics. Our model will be useful for interpreting future wide-field survey data because it is likely that understanding radio relics will require the analysis of reasonably large samples. Here, double radio relics are particularly intriguing as they form a much cleaner subsample of radio relics where the geometry and the mass ratio can be better constrained and there will be no contamination by phoenix sources. The advantage of our analytical model is that parameters can easily be varied, while full cosmological simulations can only scan a small set of cosmological and physical parameters. 

In Sec.~2 we will describe our model, including the modelling of cluster mergers, electron injection and the magnetic field. In Sec.~3 we present our results and in Sec.~4 we will conclude.

Throughout the paper, we assume a $\Lambda$CDM cosmology with $h = 0.72$, $\Omega_{\mathrm{M}} = 0.258$, $\Omega_{\mathrm{b}}=0.0441$ and  $\Omega_{\Lambda} = 0.742$. 

\section{Method}

\subsection{Cluster Mass Function}

According to the Press-Schechter formalism \citep{1974apj...187..425p}, the number of galaxy clusters of mass $[M, M+\Delta M]$ per comoving volume is given by

\begin{multline}
\label{eq:press}
  n(M, z) \,\Delta M = \sqrt{\frac{2}{\pi}} \frac{\bar{\rho}}{M}
  \frac{\delta_c(z)}{\sigma^2(M)} \left| \frac{d\sigma(M)}{dM} \right| \\
 \times \exp\!\left[ -\frac{\delta_c^2(z)}{2\sigma^2(M)} \right] \Delta M,
\end{multline}

where $\bar{\rho}$ is the mean density of the universe today,
$\delta_c(z)$ is the critical linear overdensity for a region to collapse
at redshift $z$, and $\sigma(M)$ is the current rms density
fluctuation within a sphere containing a mass $M$ given by

\be
\label{eq:sigma-mass}
  \sigma(M) = \sigma_8 \left( \frac{M}{M_8} \right)^{-\alpha},
\ee
where $\sigma_8$ is the current rms density fluctuation on
a scale of 8 Mpc, $M_8 = (4\pi/3)(8/h\, \textrm{Mpc})^3 \bar{\rho}$
is the mass contained in a sphere of radius 8 $/h$ Mpc,
and the index $\alpha = (n+3)/6$ with $n = -7/5$
is related to the primordial power spectrum. The usage of other mass functions may affect the predicted counts at large masses, but this will affect the luminosity function of relics only weakly.

Similar to \cite{cassano05} and \cite{2019ApJ...879..104L}, we build merger trees for each galaxy cluster using Monte-Carlo simulations that identify mergers, whenever the accreted mass exceeds $\Delta M = 5\times 10^{13} M_\odot$. Some relics such as the Sausage and the Toothbrush relics have been associated with major mergers with a mass ratio of $\sim 3$ \citep{2015pasj...67..114o, 2016apj...817..179j}. However, other clusters with double relics such as ZwCl0008.8+5215 and PLCK G287.0+32.9 are merging clusters with a mass ratio closer to $\sim 5$ \citep{2017arxiv171101347g, 2017apj...851...46f}. We notice that a similar approach has also been applied by \citet{2016JCAP...10..004F} to study the impact of radio emission from turbulent halos to the anomalous synchrotron background measured by Aracade2. 

At each step, the cluster mass increases from $M_1$ at time $t_1$ to $M_2$ at a time $t_2$ ($> t_1$).
The conditional probability that a halo had a progenitor of mass $[M_1, M_1 + \Delta M_1]$ at 
time $t_1$ is given by:

\be
\label{eq:prob}
  {\cal P}(\Delta S, \Delta\omega) =\frac{1}{\sqrt{2\pi}}
  \frac{\Delta\omega}{(\Delta S)^{3/2}}
  \exp \left[ -\frac{(\Delta\omega)^2}{2 \Delta S} \right] d\Delta S ,
\ee
where $\Delta S = \sigma^2(M_1+\Delta M_1)-\sigma^2(M_1)$. In our Monte-Carlo simulation we employed a time step that obeyed

\be
\label{eq:timestep}
  \Delta\omega = 0.25 \left[
    S \left| \frac{d \ln \sigma^2}{d\ln M} \right|
    \left( \frac{\Delta M_c}{M} \right) \right]^{1/2} .
\ee
The quantity $\Delta S$ can be randomly drawn from the cumulative
probability distribution of subcluster masses given by

\begin{align}
  \label{eq:subcluster}
  {\cal P}(\Delta S, \Delta\omega)
    & = \int_0^{\Delta S} {\cal P}(\Delta S', \Delta\omega) \,d\Delta S' \\
    & = 1 - \erf \left( \frac{\Delta \omega}{\sqrt{2 \Delta S}} \right) .
\end{align}
Finally, we also assumed a maximum redshift of $z_{\rm max}=2$.

\subsection{Non-thermal electron injection and evolution}

 In our simplified model, we assume that, for a given time, $t_{\rm dur}$, the CR electrons inside the relic have a set spectrum given by

\be
\label{eq:injection}
n_{\rm CR,e}(p) = C \, (p/m_e c)^{-s},
\ee
where $s$ is the index of the momentum distribution (as opposed to the energy distribution). At the shock itself, DSA injects electrons with a momentum spectrum of slope 
\be
s_{\rm inj}=2\frac{\mathcal{M}^2+1}{\mathcal{M}^2-1} ,
\ee
where $\mathcal{M}$ is the sonic Mach number \citep{1987phr...154....1b} and where we assumed an adiabatic index of $\gamma_{\rm ad} =5/3$.  
Assuming that physical conditions, such as the magnetic field, in the downstream region do not vary with distance from the shock or with time, the electron spectrum integrated over the downstream region follows a power-law with slope $s=s_{\rm inj} +1$.

The spectrum is confined to within a minimum momentum of $p_{\rm min}/m_e c = 5$ and a maximum momentum of $p_{\rm max}/m_e c = 10^4$. The values of the minimum and maximum momenta depend on the microphysics of the particle acceleration and the properties of the shock \citep{2003ApJ...583..695G}. The maximum momentum can be related to the balance between the acceleration rate and the escape rate from the acceleration region. As shown in \cite{2003ApJ...583..695G} the results will be mostly nsensitive to the choice of $p_{\rm min}$.f This is also true in our case because the assumed acceleration efficiency of electrons is a free parameter in our model. It is tuned to produce realistic radio powers, while $p_{\rm min}$ only affects the range over which the 
energy generated at shocks gets distributed. This is discussed in more detail in Appendix A.

The index of the momentum spectrum is related to the radio spectral index via $\alpha_{\nu}=(s-1)/2$, where $\alpha_{\nu}$ is the integrated spectral index within the entire downstream cooling region\footnote{The spectral index is defined as $S_{\nu}\propto \nu^{-\alpha_{\nu}}$}.

Most radio relics have integrated spectral indices in the range $\alpha_{\nu}=1.0$ to 1.5 \citep[e.g.]{2012a&arv..20...54f, 2012mnras.426...40b}. The cluster Abell 2256 is an exception in that it shows a flatter integrated spectrum, with good data available, is Abell 2256 where the spectral index is $\geq -0.9$.  \citep[e.g.][]{2008a&a...489...69b,2015A&A...575A..45T}.

The normalisation of the injection spectrum, $C$, can be determined by demanding that the CR electrons at the shock have a fixed fraction of the kinetic energy of the shock, i.e.:

\be
\label{eq:injection2}
 \int_{p_{\rm min}}^{p_{\rm max}} n_{\rm CR,e}(p)\, E(p) dp = \frac{1}{2}\eta_e \rho_{\rm g} u_{\rm sh}^2 ,
\ee
where $\eta_e$ is the acceleration efficiency, $\rho_{\rm g}$ the upstream gas density
at the shock,  and $u_{\rm shock}$ the shock speed. Formally, this is inconsistent since in DSA $\eta_e$ also depends on the minimum momentum $p_{\rm min}$. However, this dependence is model-dependent and we will treat the parameters as independent. 
Here, we assume a constant acceleration efficiency, $\eta_e=0.01$, even though there is evidence for a dependence of acceleration efficiency on Mach number. This can lead to an overestimate of relic power from low Mach number shocks \citep{2013mnras.435.1061p}. Particle-in-cells simulations have suggested that shock drift acceleration can play a role in diffusive shock acceleration which would boost the efficiency for low Mach number shocks \citep{2014apj...794..153g}. In passing, we also note that at low acceleration efficiencies we can also neglect the back-reaction on the structure of shocks.

Recent work using Particle-in-Cell simulations has suggested the potentially
significant role of Shock Drift Acceleration (DSA) in quasi-perpendicular shocks as a mechanism to inject electron from the thermal pool into the DSA regime, which should ultimately produce the radio emitting electrons probed by radio relics \citep[][]{2014apj...794..153g}. 
In SDA, in contrast to DSA, the shock itself acts as a magnetic mirror, which reflects a fraction of the incoming electrons back upstream. Simulations have shown that the minimum electron momentum required for reflection via SDA is much lower (by a factor $\sim m_{\rm e}/m_{\rm i}$) than what is commonly assumed by the 
thermal leakage model in DSA. The electron injection fraction in even low Mach number shocks can in this case be as large as $\sim 10-20\%$. Later studies by \citet{2019ApJ...876...79K} have confirmed the above picture for the minimum injection momentum required for electrons to be accelerated by SDA, however they have questioned whether SDA is efficient enough to inject such electrons into the DSA regime.

Unlike in \cite{2003ApJ...583..695G}, we only consider the shock acceleration of electrons and ignore protons, motivated by the fact that no diffuse $\gamma$-ray emission of hadronic origin has been detected by FERMI-LAT, which places very low limits on the cosmic-ray proton contents of the ICM \citep[][]{2016apj...819..149a}.

We assume that the density of the ICM, $\rho_{\rm g}$, is given by a $\beta$-model, even if that may be a poor approximation to the density distribution in merging galaxy clusters, i.e.

\be
  \label{eq:beta}
  \rho_{\rm g}(r) = \rho_{\rm g}(0) \left[1 + (r / r_{\rm c})^2 \right]^{-3\beta/2},
\ee
where $r_{\rm c}$ is the core radius and we set the parameter $\beta = 2/3$. For the core radius, we adopt $r_{\rm c} = 0.1 \,R_{200}$ \citep[e.g.,][]{2003mnras.340..989s}. The central gas density $\rho_g(0)$ can be derived from the total gas mass via $M_{\rm g} = f_{\rm g} M_{200}$. The last parameter in Eq.~(\ref{eq:injection2}) is the shock speed, $u_{\rm shock}$, but this we will  discuss in the next section.\\

Mildly relativistic electrons, which may have long radiative life times in the cluster periphery, may also be re-energized by merger shocks, leading them to reach the relativistic energies required to produce the nonthermal radiation. However, here we do not explicitly consider re-acceleration of pre-existing populations of non-thermal electrons nor additional re-acceleration of electrons (e.g. by Fermi-II processes) in the downstream region of shocks \citep{2017apj...840...42k}. However, the fiducial value of our acceleration efficiency can effectively only be explained if pre-existing populations of electrons are re-accelerated. 
Re-acceleration is believed to generate particle spectra that are flatter than those of the seed particles, at least if the injection spectrum relative to the last shock is flatter than the energy spectrum attained by fossil particles.  
Some recent works \citep{2013mnras.435.1061p, va14relics, 2017apj...840...42k, 2019arXiv190700966B} suggest that pre-existing populations may be necessary to explain the acceleration efficiency, so this will be an avenue for further study.
Pre-existing populations of relativistic electrons lead to enhanced shock acceleration that can be modelled using an effectively higher acceleration efficiency that depends on the shock Mach number, as shown by numerical simulations \citep[][]{ka12,2013mnras.435.1061p}.

\subsection{Merger shock properties}

Since the work of \citet{2003ApJ...583..695G} there has been substantial progress in our understanding of cluster outskirts, particularly through numerical simulations and deep X-ray observations \citep[e.g.]{2013SSRv..177..195R}.

Cluster merger shocks have been studied with numerical methods \citep[e.g.]{1999ApJ...520..514T, 2001ApJ...561..621R}. Most shocks within self-gravitating systems are weak, as to a first approximation the Mach number is of the order of the ratio between the infall velocity and the circular velocity of halos, which gives $\mathcal{M} \sim \sqrt{2}$ \citep{2003ApJ...583..695G}.

\cite{2012mnras.421.1868v} have shown why radio relics sit primarily on the outskirts of galaxy clusters.
The strength of a spherical shock wave that travels outwards varies due to (i) geometry ($\propto r^{-2}$), (ii) the gas density profile and also possible dissipative processes.
In galaxy clusters, the gas density profiles are fairly flat in the centre, and steepen to a slope of around $r^{-2.5}$ around $R_{500}$. 

\cite{2019arXiv190403052Z} have shown that, unlike in a homogeneous medium, the steep gas density gradient in cluster outskirts, the merger shock maintain its strength over a long distance.
Backed up by numerical simulations, \cite{2019arXiv190403052Z} have found a "habitable zone" of merger shocks, where shock waves maintain their strength for a relatively long time. 

\cite{2018apj...857...26h} called the merger shocks that produce relics, "axial shocks", as opposed to the tangential shocks that propagate perpendicular to the merger axis and before the core passage of Dark Matter centres. The axial shocks are found to emerge about 1 Mpc from the X-ray peaks. As others, \cite{2018apj...857...26h} observed that the shock ahead of more massive halo is stronger than that ahead of lower mass halo \citep{2017apj...841...46m}.

In hydrodynamical simulations of cluster mergers \cite{2018apj...857...26h} found that, as the shocks propagate outward, the Mach number increases while the shock speed stays fairly constant. The flux of kinetic energy through the shock surfaces slowly decrease with time since the gas density decreases going outward. The kinetic energy flux through the shock is found to peak at about 1 Gyr after the shocks are launched, at distances of  $\sim 1-2$ Mpc from the cluster center. At that time, the shocks have Mach numbers of $\mathcal{M}\sim 2-3$. 

\begin{figure}
\center{\includegraphics[width=\columnwidth]{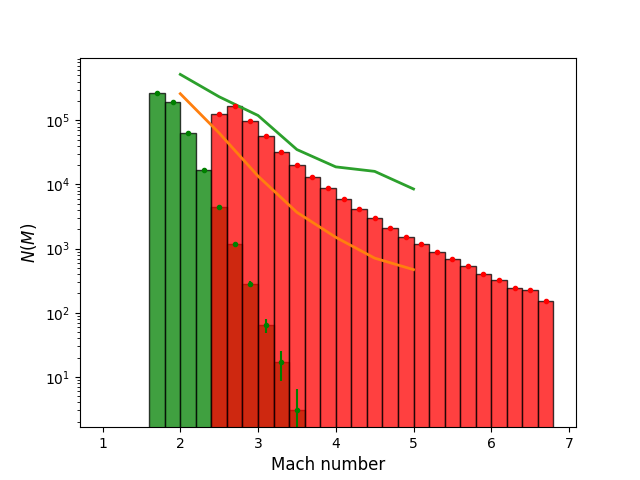}}
\caption{The distribution of Mach numbers for a simulation with our fiducial set of parameters. The green histogram shows the Mach numbers of the shocks in the more massive cluster and the red one the Mach number in the less massive cluster. The lines show the values for Mach numbers found in a large set of hydrodynamical simulations \citep[][]{va14relics}.}
\label{fig:mach}
\end{figure}

Based on these hydrodynamical simulations, we assume that shocks in the more massive of the two clusters (see \citealt{2018apj...857...26h}) propagate with constant velocity, $u_{\rm sh}$, according to 

\be
r_{\rm sh}= 0.5 \cdot R_{200} + u_{\rm sh} \cdot t ,
\label{eq:radius}
\ee
with

\be
\label{eq:velshock}
u_{\rm sh, main} = 1,500\,\mathrm{km/s}\cdot(1+\xi)^{1/2} \left ( \frac{M}{5\times 10^{14}M_\odot} \right )^{1/2} ,
\ee
where $\xi=M_2/M_1$. The velocity of the shock front in the subcluster is given by

\be
\label{eq:machsub}
u_{\rm sh, sub} = u_{\rm sh, main}\cdot ˜\xi^{-3/4} ,
\ee
as given in \cite{1999ApJ...520..514T}. The Mach numbers are then given by dividing by the sound speed.
Unlike in previous studies, we do not assume an isothermal ICM in order to calculate the Mach numbers of the shocks. X-ray observations have shown that there is a general trend for the temperature of the ICM to decline with distance from the centre.  Here, we assume that the cluster temperature decreases with radius according to $T/T_{\rm vir}=1.19-0.74 R/R_{\rm vir}$ as inferred from X-ray observations of nearby galaxy clusters \citep{2007a&a...461...71p}, where $T_{\rm vir}$ is the virial temperature of the cluster or subcluster. The resulting Mach number distribution, shown in Fig.~\ref{fig:mach}, agrees with the distribution found in cosmological simulations \citep{2010NewA...15..695V}.
While the temperature distribution with radius in merging cluster will differ substantially from what are commonly called isolated and relaxed clusters, the regions into which merger shocks propagate is believed to be very close to that of non-merging clusters.  

In order to calculate the life time of the relic, we assume that the relic emits for a time $t_{\rm dur}=r_c/v_i$, where $r_c$ is the core radius of the main cluster (which we take to be $r_c=0.2 R_{\rm vir}$) and $v_i$ is the impact velocity of two merging clusters of mass $M_1$ and $M_2<M_1$, which is given by

\be
v_i=\left [\frac{2G(M_1+M_2)}{r_1} \left ( 1-\frac{1}{\eta_v}\right ) \right ]^{1/2} ,
\label{eq:vel}
\ee
with $\eta_v=4(1+M_2/M_1)^{1/3}$ \citep[][]{sarazin2002}.

\subsection{Magnetic field}

\label{sec:magnetic}
The magnetic field is another great uncertainty in calculating the radio power from radio relics.  Historically, most observations of radio relics suggest fairly high fields of 2-7 $\mu$G, also based on the lack of detection in the associated Inverse Compton emission from the same region, under simplistic assumptions \citep{2019SSRv..215...16V, 2010apj...715.1143f, 2009pasj...61..339n}.

More recently, the width of the radio relics has been used to constrain the magnetic field strength in the relic. If we know the bulk speed of the gas motion behind the shock that causes the relic, we can calculate the width given the characteristic timescale of the electron energy losses \citep{2010sci...330..347v}. One of the best studied magnetic fields in relics is that of the Toothbrush relic. Detailed observations and modelling of the Toothbrush relic have suggested that the magnetic field should be $<5\,\mu$G, with almost constant strength on $\sim \rm Mpc$ scales \citep{2018apj...852...65r,2019arXiv191108904R}.

Here we test two scenarios: (i) the magnetic field is proportional to the thermal energy density in the ICM at the shock location,
$\epsilon_B = B^2/(8\pi) \simeq 0.015 \epsilon_{\rm th}$, and (ii) the magnetic field in relics is a constant 2 $\mu$G. The thermal energy density inside a cluster is computed via $\epsilon_{\rm th}=(3/2)n_{\rm th}kT_{\rm ICM}$ and
\be
n_{\rm th}=\frac{3f_{\rm g}M_{\rm vir}}{4\pi\mu m_u R_{\rm vir}^3} ,
\ee
with $T_{\rm ICM} = \mu m_u GM_{\rm vir}/(2 R_{\rm vir})$ given by the virial theorem.

The former magnetic field estimate presents usually a lower limit to the magnetic field strengths in clusters and 2 $\mu$G may be a high value for a volume-filling magnetic field inside the relic regions. See also \cite{2016mnras.462.2014d} and \citet{2019arXiv190911329W} for modelling  of the magnetic field evolution in giant radio relics.

\subsection{Specific radio power}

Once we have calculated the number density of non-thermal electrons $n_e(\gamma)$, we find the
synchrotron emissivity at frequency, $\nu$, via
\be
\label{eq:sync}
  J_{\nu} = \frac{\sqrt{3} \, e^3 B}{m_{\rm e} c^2}
    \int_{\gamma_{\rm min}}^{\gamma_{\rm max}} \int_0^{\pi/2}
    n_{\rm CR, e}(\gamma, t) F(\nu/\nu_c) \sin^2 \,\theta \,d\theta \,d\gamma ,
\ee
where $\theta$ is the pitch angle of electrons with respect to the magnetic
field, $\nu_c = (3/2) \,\gamma^2 e B \sin\theta /(2\pi m_{\rm e} c)$ is the
critical frequency and $F(x)$ is the usual synchrotron kernel \citep{1979rpa..book.....r}.

Integrating the emissivity over the volume of the relic yields the specific power of the radio relic.  The volume of the relic is given by a fixed segment of a spherical shell, whose thickness is set by the cooling length of the electrons (more details below, see also Eq.~\ref{eq:coollength}).

Hence, we set 
\be
\label{eq:volume}
V = \Omega r_{\rm shock}^2 l_{\rm cool} ,
\ee
where $\Omega$ is the solid angle subtended by the relic with the cluster centre as apex. We use a fiducial value of $\Omega=\pi/5$, which corresponds to an opening angle of $25^\circ$.  Finally, $r_{\rm shock}$ is the cluster-centric distance of the shock given in Sec. 2.3. The cluster-centric relic distance, $r_{\rm shock}$ is given by Eq.~(\ref{eq:radius}). Finally, $l_{\rm cool}$ is the cooling length of the electrons at the given frequency given by \cite{2017apj...840...42k}:

\be
\label{eq:coollength}
l_{\rm cool} = 120\,\mathrm{kpc}\, \kappa \left ( \frac{u_{\rm d}}{10^3\mathrm{km/s}}\right ) \left [ \frac{\nu(1+z)}{0.61 \mathrm{GHz}}\right ]^{-1/2} ,
\ee
where $u_{\rm d}$ is the downstream velocity, the factor $\kappa$ depends on the downstream magnetic field, $B_{\rm d}$ (expressed in $\mu\mathrm{G}$) as

\be
\kappa=\frac{(5 \mu\mathrm{G})^2}{B_{\rm d}^2+B^2_{\rm CMB}(z)}\left ( \frac{B_{\rm d}}{5 \mu\mathrm{G}}\right )^{1/2} .
\ee
In this expression, $B_{\rm CMB}(z)=3.24\,\mu\mathrm{G}(1+z)^2$ is the equivalent field strength of the CMB radiation field. Since we are considering binary mergers, we always have double relics, and the relative sizes and powers of the individual relics is determined by the respective virial radius, $R_{\rm vir}$, of the subcluster.

The downstream magnetic field is modelled as described in Sec.~\ref{sec:magnetic} and, for simplicity, the downstream speed is taken to be $u_{\rm d}=u_{\rm shock}$.

Finally, the flux density at the same frequency, including the $K$-correction, is given by
\be
  \label{eq:flux}
  S_{\nu} = \frac{(1+z) P_{\nu(1+z)}}{4\pi D_{\rm L}^2(z)} ,
\ee
where $D_{\rm L}(z)$ denotes the luminosity distance.

Most relics are observed to be in clusters whose merger axis is close to the plane of the sky. \cite{2018ApJ...862..160W} constrained the angle between the line-of-sight and the merger axis by means of a comparison with simulations of cluster mergers. Within the 68\% confidence interval, they found that the merger axis lies within 53 degrees of the plane of the sky.
Hence, we draw the merger axis from an isotropic random distribution and require that the merger axis lies within 30 degrees to the plane of the sky. A table of the main parameters of our model can be found in Tab.~\ref{tab:parameters}. All software is written in {\tt python} and a typical run for a region of 100 $\times$ 100 degrees requires $< 30$ minutes to complete. 

\begin{table}
    \centering
     \caption{Main model parameters.}
    \begin{tabular}{@{}lll@{}}
     \hline
    Parameter & meaning & fiducial value\\
    \hline
    $p_{\rm min}$ & Minimum momentum factor of CRe & $5\, m_e c$\\
    $p_{\rm max}$ & Maximum momentum factor of CRe & $10^4 m_e c$\\
    $M_{\rm min}$ & Minimum halo mass & $5\cdot 10^{13}M_\odot$ \\
    $M_{\rm max}$ & Maximum halo mass & $5\times 10^{15}M_\odot$ \\
    $z_{\rm max}$ & Maximum redshift & 2 \\
    $\Delta M$ & Minimum mass growth for merger & $10^{13}M_\odot$  \\
    $\sigma_8$ & rms fluc on 8 Mpc scale & 0.81 \\
    $\delta_{\rm sky}$ & maximal angle of merger axis with sky & $30^\circ$\\
    $\Omega$ & Solid angle of relics & $\pi /5$\\
    $f_g$ & gas mass fraction & 0.17 \\
    $\eta_e$ & acceleration efficiency & 0.01 \\
    $\alpha$ & index of primordial power spectrum & 4/15 \\
    $\beta$ & index for ICM distribution & 2/3 \\
    \hline
    \end{tabular}
    \label{tab:parameters}
\end{table}

\section{Results}

\begin{figure}
\center{\includegraphics[width=\columnwidth]{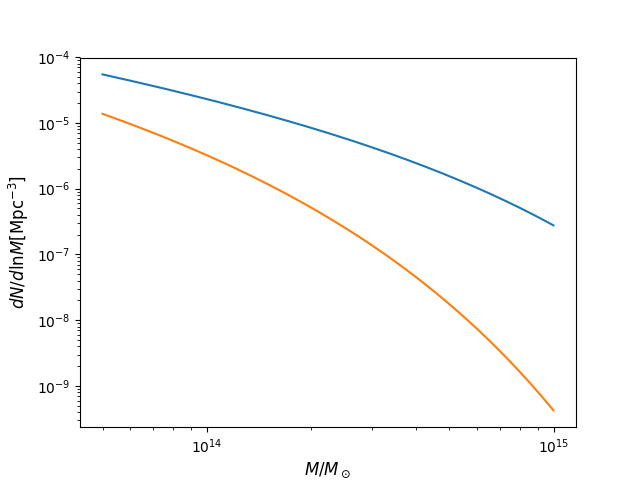}}
\caption{The mass function of the simulated galaxy clusters for $z=1$ (orange) and $z=0$ (blue). See text for details.}
\label{fig:massfctn}
\end{figure}

In the following, we show source statistics as well as artificial sky maps for an area of sky of 100 $\times$ 100 degrees. This sky area is big enough to contain the effects of cosmic variance and can then be scaled to any survey area of interest. In the assumed cosmology and with $z_{\rm max}=2$ this contains a volume of about 14.1 Gpc$^3$.
Before we look at the distributions of radio luminosities and fluxes, we show in Fig.~\ref{fig:massfctn} the halo mass function at redshifts  $z=1$ and $z=0$, as they are predicted from the Press-Schechter formalism. In total, there are 744,013 clusters in this area and up to the maximum redshift of $z_{\rm max}=2$ assuming a halo minimum mass (Dark Matter) of $4.15\times 10^{13}M_\odot$. It should be noted that our results on radio relics are not very sensitive to the choice of the maximum redshift as most of detectable relics are located at $z \leq 0.2$.

\begin{figure}
\center{\includegraphics[width=\columnwidth]{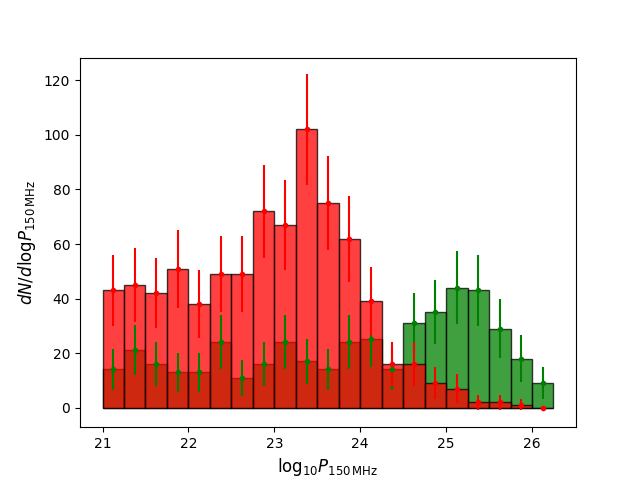}}
\caption{Distribution of powers at 150 MHz (in units of W/Hz) for our fiducial set of parameters. The green bars represent the relics in the main clusters and red the relic in the sub-clusters.}
\label{fig:power150}
\end{figure}

\begin{figure}
\center{\includegraphics[width=\columnwidth]{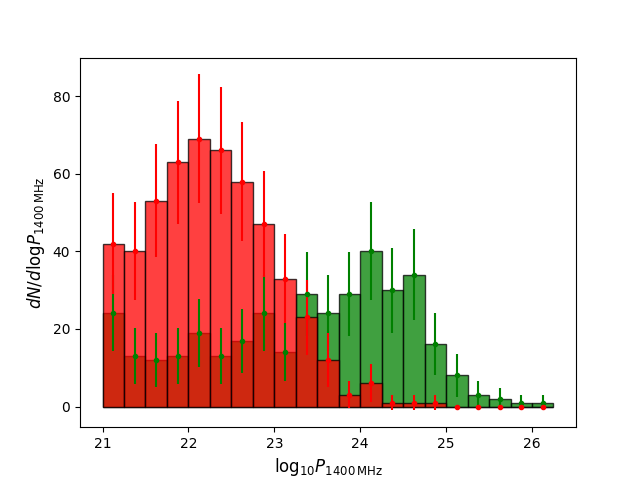}}
\caption{Distribution of powers at 1.4 GHz (in units of W/Hz) for our fiducial set of parameters. The green bars represent the relics in the main clusters and red the relic in the sub-clusters.}
\label{fig:power1400}
\end{figure}

For our fiducial set of parameters, Fig.~\ref{fig:power150} and \ref{fig:power1400} show histograms of the specific radio power at 150 MHz and 1.4 GHz, respectively. The green bars represent the relics formed in the main clusters and red the relics in the sub-clusters. The main clusters produce the brighter relics, in line with what is generally found in observations. Even though the Mach number is higher in the sub-clusters (owing to the lower sound speed there), the relics are bigger and injection goes on for longer in the main clusters. The distribution at the different frequencies is as expected given the relics' spectral indices. The peak of the distribution is mostly affected by our assumption of the shock dynamics, i.e. the Mach number distribution. The total number of relics depends on a number of factors, such as the acceleration efficiency, the size of the relics as well as the magnetic fields. Some of the dependencies we will test below.

\begin{figure}
\center{\includegraphics[width=\columnwidth]{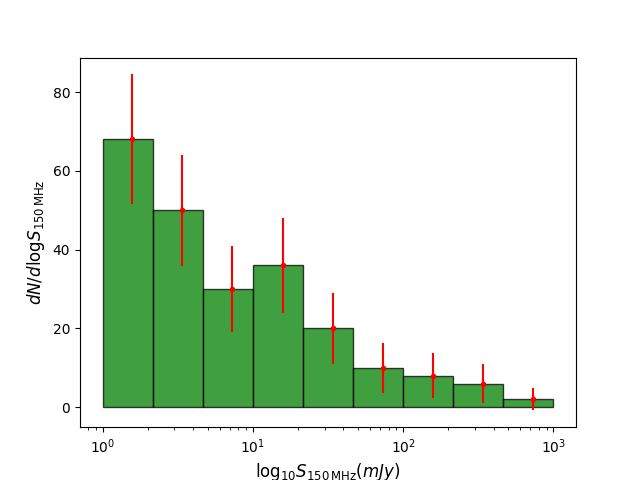}}
\caption{Flux distribution at 150 MHz for our fiducial set of parameters. Here double relics from binary mergers are counted individually.}
\label{fig:flux150}
\end{figure}

\begin{figure}
\center{\includegraphics[width=\columnwidth]{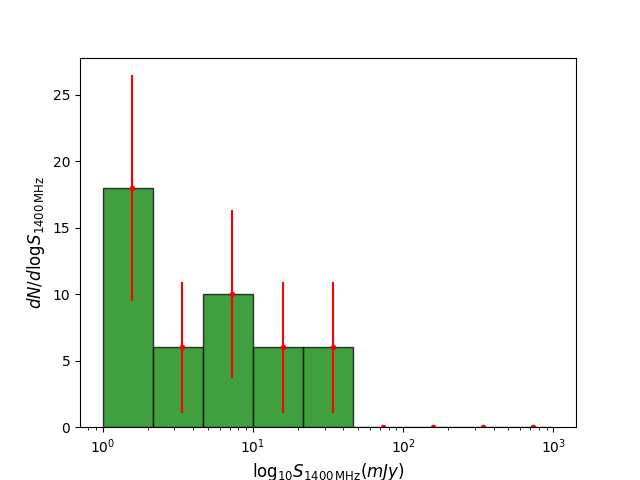}}
\caption{Flux distribution at 1.4 GHz.  Here double relics from binary mergers are counted individually.}
\label{fig:flux1400}
\end{figure}

More readily accessible are the distribution of fluxes that are shown next. Fig.~\ref{fig:flux150} and \ref{fig:flux1400} show histograms of radio relic fluxes at 150 MHz and 1.4 GHz, respectively. Note here that in these histograms a double relic would count as two relics.  The lowest flux densities of known relics lie at around 2 mJy. A prediction of our model is the emergence of a large population of faint radio relics only visible at low flux levels (e.g. $\leq 10$ $\rm mJy$ at 150 MHz), which corresponds to shocks launched in the least massive cluster of colliding pairs.

\begin{figure}
\center{\includegraphics[width=\columnwidth]{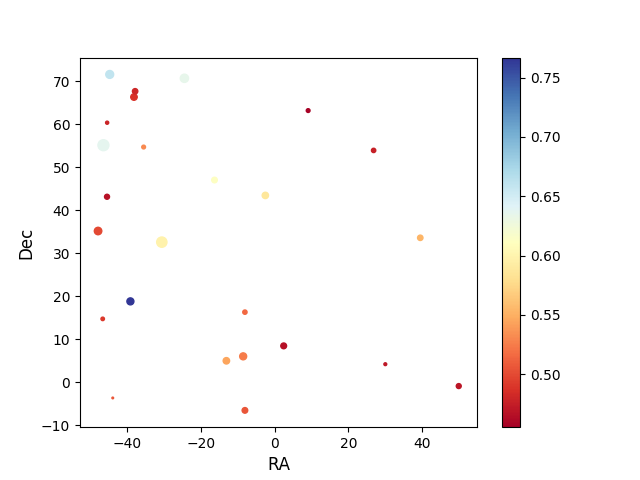}}
\caption{Map at 1.4 GHz for our fiducial parameters and equipartition magnetic fields. The colours of the circles denotes the magnetic field strength (in $\mu$G) and the sizes of the symbols are proportional to the fluxes. Here the relic emission from main and subclusters are added.}
\label{fig:map1400}
\end{figure}

\begin{figure}
\center{\includegraphics[width=\columnwidth]{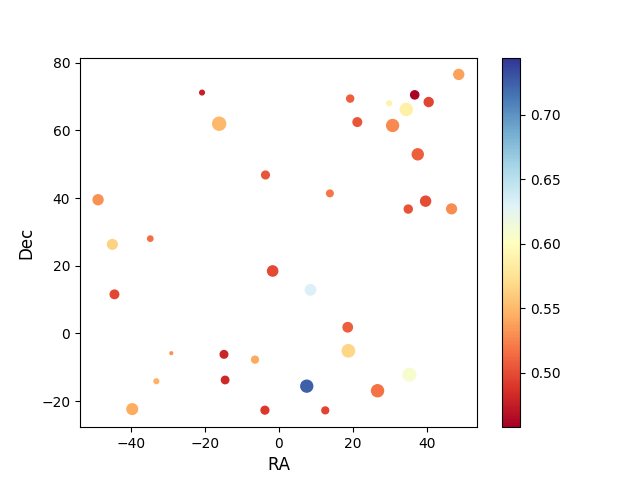}}
\caption{Map at 150 MHz for our fiducial parameters and equipartition magnetic fields. The colours of the circles denotes the magnetic field strength (in $\mu$G) and the sizes of the symbols are proportional to the fluxes. As in Fig.~\ref{fig:map1400} the relic emission from main and subclusters are added.}
\label{fig:map150}
\end{figure}

In Fig.~\ref{fig:map1400} and \ref{fig:map150} we show maps of the relic positions in the 100 $\times$ 100 degree field at 150 MHz and 1.4 GHz, respectively. The choices for the limits in right ascension and declination are obviously arbitrary. The colours of the circles denote the magnetic field strength and the sizes of the symbols are proportional to the fluxes at the given frequency.

\begin{figure}
\center{\includegraphics[width=\columnwidth]{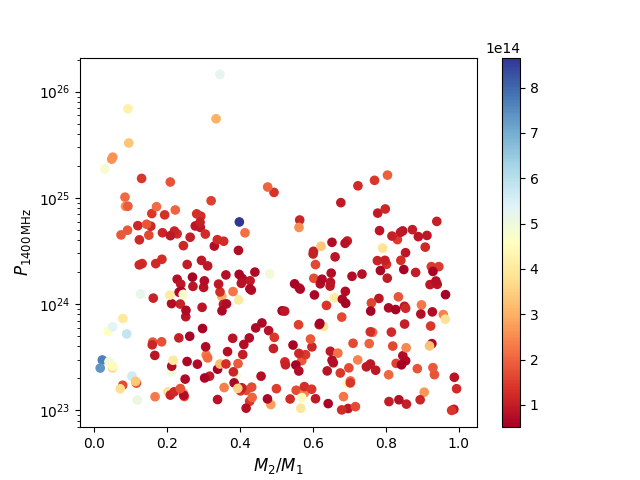}}
\caption{Power in units of W/Hz versus mass ratio of the merger, $M_2/M_1$, where $M_2<M_1$, for our fiducial model and equipartition magnetic fields. The colour denotes the total mass, $M_1+M_2$.}
\label{fig:M1M2}
\end{figure}

In Fig.~\ref{fig:M1M2} we test the dependence of the relic power on the mass ratio of the merger for our fiducial set of parameters. 
The power versus mass ratio shows no clear trends. The only trend that emerges from Fig.~\ref{fig:M1M2} is that high powers at small ratios only occur for high-mass clusters (represented by the blue dots). It remains to be seen whether future samples of radio relics confirm this.

\begin{figure}
\center{\includegraphics[width=\columnwidth]{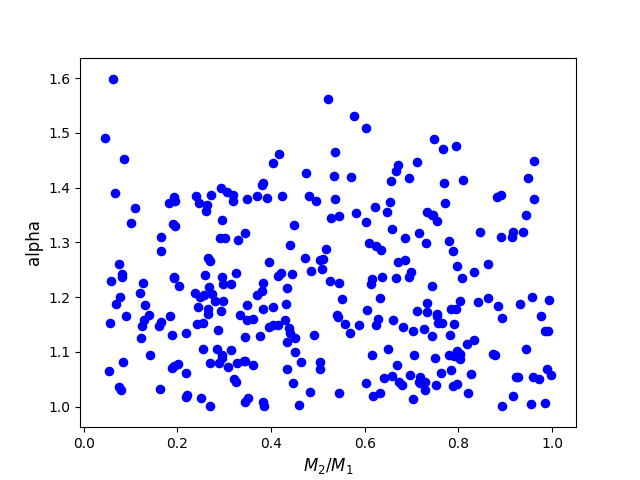}}
\caption{Radio spectral index versus mass ratio of the merger, $M_2/M_1$ (where $M_2<M_1$), for our fiducial model and equipartition magnetic fields.}
\label{fig:alpha}
\end{figure}

\begin{figure}
\center{\includegraphics[width=\columnwidth]{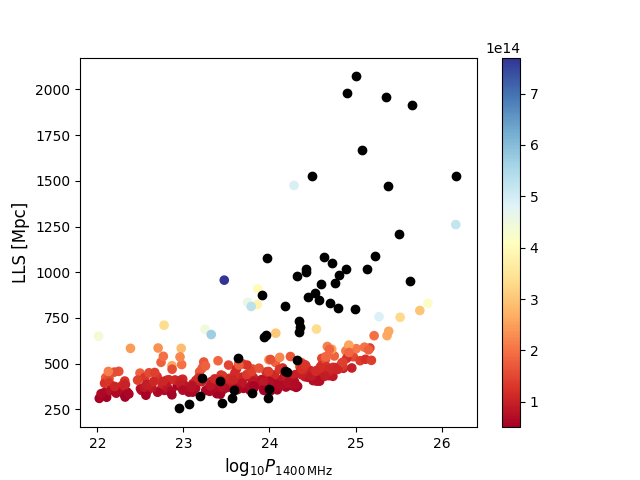}}
\caption{Largest linear size (LLS) in Mpc (physical - not comoving) versus total power of relics for our fiducial model and equipartition magnetic fields. The colour codes show the radio spectral index. The black dots represent observed values for relics in a sample compiled in Nuza et al. (2017).}
\label{fig:LLS}
\end{figure}

The same absence of a general trend is seen in the radio spectral index, $\alpha$ as a function of mass ratio (see Fig.~\ref{fig:alpha}). However, the range of spectral indices lies within the range of observed relics, not considering phoenix-type relics. In Fig.~\ref{fig:LLS}, we are plotting the largest linear size (LLS) of the relics as a function of their power at 1.4 GHz. The colour-coding represents the total mass of the host cluster. We see that the LLS grows with the power of the relic as expected. Moreover, the largest relics all occur in very massive clusters. With black dots we overplot the observed values for relics in a sample compiled in \cite{2017mnras.470..240n}. The observed LLS depends a bit more strongly on the power than predicted by our fiducial model, yet the observed values are within the range of our model.

\begin{figure}
\center{\includegraphics[width=\columnwidth]{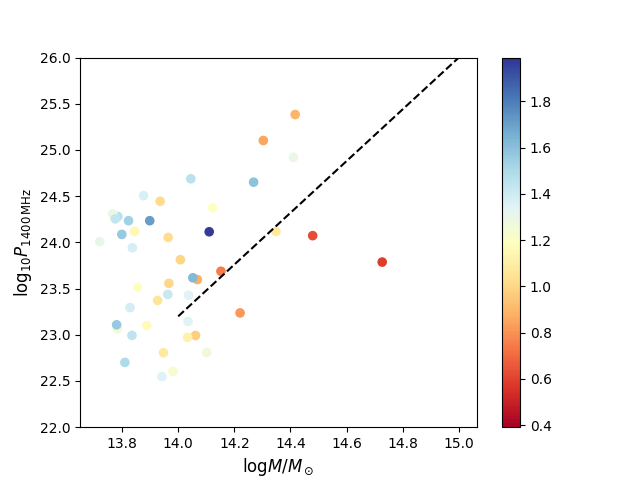}}
\caption{$P_{1400}-M$ plot for all relics that have a flux of $>$ 1 mJy in our fiducial model and equipartition magnetic fields. The colour of the dots corresponds to the redshift of the host cluster. The specific power is expressed in units of W/Hz. The black line indicates a slope of 2.8 as found by De Gasperin et al. (2014).}
\label{fig:powermasscompare}
\end{figure}

In Fig.~\ref{fig:powermasscompare}, we plot the $P_{1400}-M$ relation for our fiducial model and all relics that have a flux of $>$ 1 mJy. Most interestingly, the $P_{1400}-M$ distribution follows roughly the observed slope of 2.8 as found by \cite{2014mnras.444.3130d}. Colour-coded is the redshift of the host clusters. The distribution of all relics in our sample peaks at a redshift of about 0.75, however neglecting the higher inverse Compton losses at the higher redshifts. In addition, our model predicts many more relics at lower powers that will have been missed by current observations.  The lack of a clearly defined sample of clusters does not allow any further comparisons to observations at this point.

Below we show results when two key parameters of our model are varied: the magnetic field strength as well as the life time of the relics. These have emerged as the key parameters. Other, fairly poorly constrained parameters could also be varied, such as the solid angle that the relics subtend with respect to the cluster centres, or the acceleration efficiency, $\eta_e$. Higher efficiencies shift the peak power (and the flux) to higher values and thus increase the number of detectable relics. However, the latter parameters can be scaled fairly straightforwardly to our fiducial model. 

\subsection{Varying the magnetic field strength}

\begin{figure}
\center{\includegraphics[width=\columnwidth]{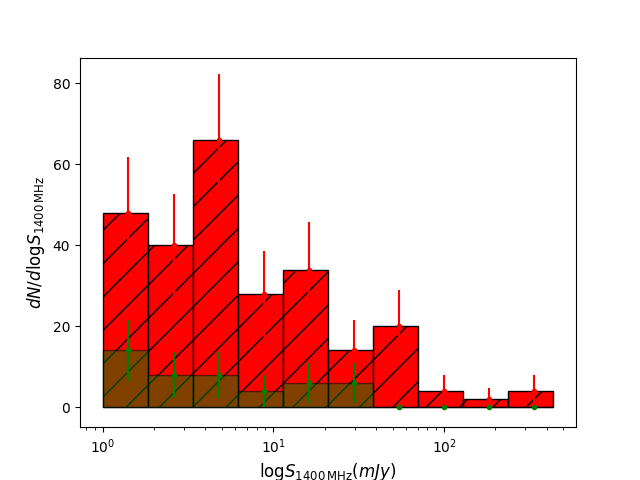}}
\caption{Flux distribution at 1.4 GHz for. The green (hatched) histogram is for $B=2\,\mu$G and red for an equipartition field.}
\label{fig:fluxvarB}
\end{figure}

\begin{figure}
\center{\includegraphics[width=\columnwidth]{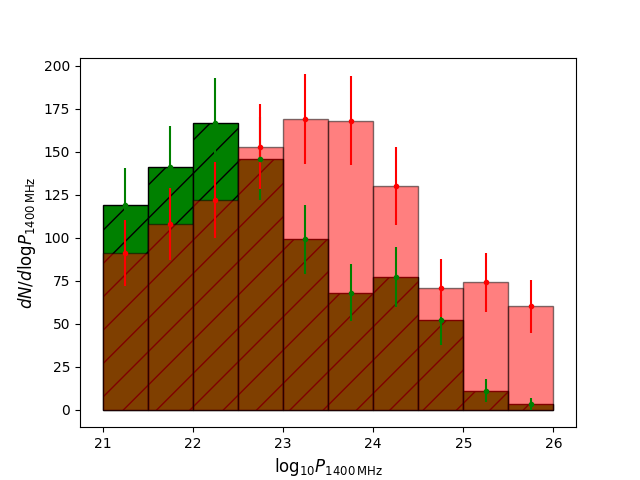}}
\caption{Distribution of powers at 1.4 GHz for. The green (hatched) histogram is for $B=2\,\mu$G and red for an equipartition field.}
\label{fig:powervarB}
\end{figure}

In Fig.~\ref{fig:fluxvarB} we show the flux distribution at 1.4 GHz for the two magnetic field strengths probed by our models. The red histogram is for $B=2\,\mu$G and green for a field that is proportional to the thermal energy density as described in Sec.~\ref{sec:magnetic}. A constant field of $B=2\,\mu$G yields a substantially larger number of relics that also go to higher flux densities. The distribution of powers for the two magnetic field models look very similar and the magnetic field only shifts the peak of the distribution (to higher powers for higher fields - see Fig.~\ref{fig:powervarB}).  A volume-filling field of $2\,\mu$G is higher than the mean equipartition field that we find. In case that relics have such high volume-filling fields, this would favour a lower acceleration efficiency.

\subsection{Varying the duration of the relics $t_{\rm dur}$}

Next, we varied the duration during which the relic is in an active state, i.e. where the acceleration of particles at the shock produces the spectrum given by Eq.~(\ref{eq:injection}). After that time, the relic emission stops.

\begin{figure}
\center{\includegraphics[width=\columnwidth]{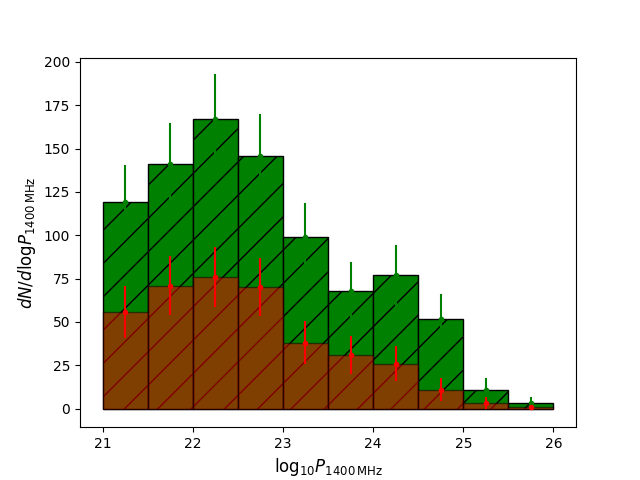}}
\caption{Power distribution at 1.4 GHz for  $t_{\rm dur}=1.5\, r_c/v_i$ (red) and $t_{\rm dur}=2\, r_c/v_i$ (green) in units of W/Hz.}
\label{fig:powervart}
\end{figure}

In Fig.~\ref{fig:powervart}, we show the power distributions at 1.4 GHz for $t_{\rm dur}=1.5\, r_c/v_i$ and $t_{\rm dur}=2\, r_c/v_i$, respectively. As is clear from the plot, the life time has a large impact on the number statistics of relics. 

By decreasing the life time by 25\%, the number of relics in the flux range shown in Fig.~\ref{fig:powervart} drops by a factor of more than 2. A longer life time leads to more relics because they are visible for longer and they propagate out to larger radii and thus fill larger volumes, leading to larger luminosities.

\subsection{Cosmology}

\begin{figure}
\center{\includegraphics[width=\columnwidth]{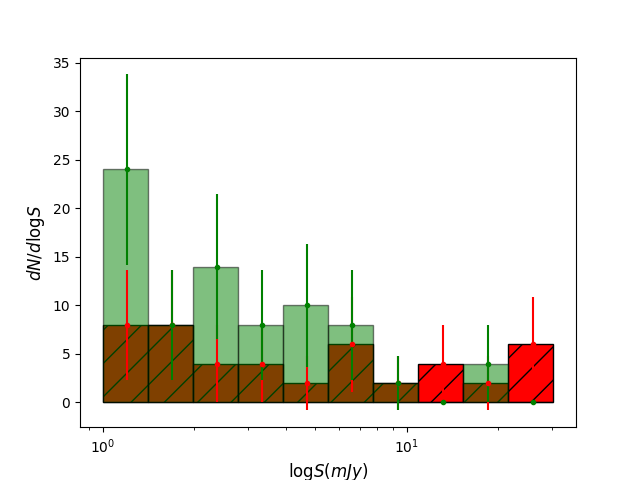}}
\caption{Flux distribution at 1.4 GHz for $\sigma_8=0.81$ (red, hatched) and $\sigma_8=0.9$ (green).}
\label{fig:fluxsigma}
\end{figure}

\begin{figure}
\center{\includegraphics[width=\columnwidth]{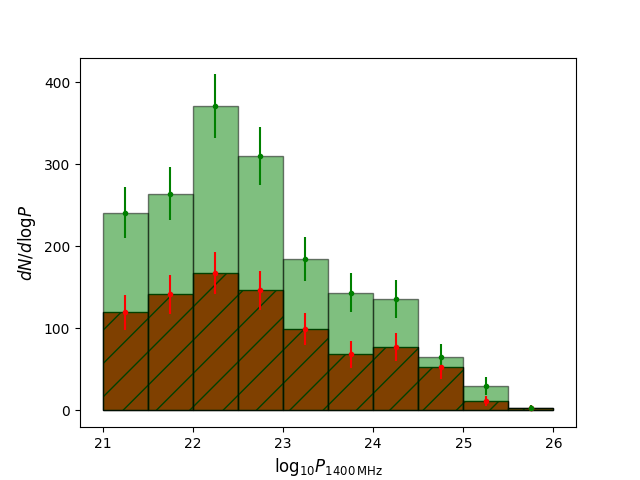}}
\caption{Power distribution at 1.4 GHz for $\sigma_8=0.81$ (red, hatched) and $\sigma_8=0.9$ (green) in units of W/Hz. }
\label{fig:powersigma}
\end{figure}

The impact of cosmological parameters on the expected statistics of radio relics have never been explored in theoretical work, despite the evidence that a change in $\sigma_8$ affects the distribution of energy dissipated in shocks. This has been found in cosmological simulations \citep[][]{va09shocks}. Our analytical approach allows us to easily explore the relation between radio relic emission and cosmology, which would otherwise be a demanding numerical task. 
Given the large uncertainties in the particle acceleration efficiency and shock dynamics, double radio relics are, at least at this stage, no suitable cosmological probes. Nonetheless, we have compared in Fig.~\ref{fig:fluxsigma} the flux distribution for our fiducial value of $\sigma_8=0.81$ to a value of $\sigma_8=0.9$, which is far beyond its current uncertainty \citep[][]{2016A&A...594A..13P}. For completeness, we also compare the luminosities in Fig.~\ref{fig:powersigma}. A larger $\sigma_8$ is found to produce a slightly more extended tail of relics with a high power, which can be understood on the basis that the merger rate increases with increasing $\sigma_8$. Owing to the fact that only rare merger events determine the tail of the power distribution, producing a large statistics of such events requires simulations in large cosmic volumes at high spatial resolution.  Differences related to cosmology and rare merger events, which are also relevant to explain powerful relics at high redshift (e.g. \citealt{2016mnras.463.1534b}) can be efficiently studied with our semi-analytical approach.

\section{Conclusions}

We have presented the results of a semi-analytical framework to model the evolution and power distribution of radio relics. While studying the role of such parameters with cosmological (magneto)-hydrodynamic simulations is difficult and costly, our semi-analytical code allows us explore the parameter space that is associated with radio relics.

The properties of merger shocks, at least based on cosmological simulations, are expected to be fairly independent of the host cluster mass and on the merger configuration, the statistical properties of the population of radio relics are strongly dependent on key, and largely unconstrained, physical parameters related to the propagation of merger shock waves in the intracluster medium: namely the acceleration efficiency, the magnetic field strength at the relic, the geometry of the relic and the duration of the electron acceleration at merger shocks. 

Current surveys do not yet lend themselves to a thorough comparison with models. The NVSS that surveys the entire Northern sky above a declination of $-40^\circ$ only has 12 double radio relics identified \citep{2017mnras.470..240n}. These relics have flux densities ranging from 9 to 104 mJy (for both components). Future surveys that have a better sensitivity to extended, low-surface brightness emission will provide a bias-free census of double relics.

Forthcoming surveys (such as LoTSS, \citealt{2019a&a...622a...1s, 2013pasa...30...20n}, the LOFAR low-band survey LOL, De Gasperin et al. in prep.), EMU \citealt{2011pasa...28..215n}) will yield large samples of radio relics and statistical studies will rapidly gain importance. Finally, one may add that
one of the contaminating foregrounds of the redshifted 21 cm signal from the Epoch of Re-ionization (EoR) are extragalactic diffuse radio sources, such as radio halos and relics. For radio halos the contribution to the signal predicted for the Square Kilometre Array (SKA) has been studied by \cite{2019ApJ...879..104L}.

There have been hydrodynamical simulations of binary cluster mergers but, again, the parameter space that has been explored is small. However, the model presented here can easily accommodate any parametrisation of the shock dynamics as long as it can be associated with the underlying dynamics of the DM halos.

The main strengths of our model are:

1. Our models reproduces the Mach number distributions that is found in cosmological hydrodynamical simulations.

2. Our models yields a plausible number of relics for reasonable parameters, such as the acceleration efficiency that has been used in \cite{2008MNRAS.391.1511H}. Moreover, we predict a large number of relics below current detection limits, which will be testable with future radio surveys.

3. Many authors have remarked on the fact that most merger-related, giant shock waves have Mach numbers between 1 and 2. As a result, standard DSA arguments imply that the spectra of particles accelerated in these shock waves are steep, and hence of minor importance for explaining non-thermal phenomena in clusters. \cite{2003ApJ...583..695G} conclude that efficient particle acceleration occurs mainly in minor mergers, i.e. mergers between clusters with very different masses. Our results agree that mergers of smaller clusters onto very large clusters can cause bright radio relics. However, the distribution of radio powers with $M_2/M_1$ is fairly flat between $0.2<M_2/M_1<1$ and the radio luminosity is mainly a function of total cluster mass, as indicated by observations (see Fig.~\ref{fig:M1M2}).

4. We reproduce a slope of the $M-P_{1400}$ relation that agrees with observations, albeit large observational uncertainties. The lack of a clearly defined sample of clusters does not allow any comparisons to observations at this point.

The main weakness of our approach is the large uncertainty in our input parameters. We find that the largest uncertainties are related to conditions for particle acceleration as well as the detailed dynamics of the shocks. In joint X-ray and radio observations of relics, the particle acceleration efficiency is reasonably constrained but the parameter space is constrained by the small available sample size  \citep[e.g.][]{2017mnras.470..240n,2019arXiv190700966B}. Another shortcoming is that we are not including the fading of relics after injection by computing radiative losses as well as re-acceleration.  Moreover, our procedure cannot properly model rare triple merger configurations, which are believed to be responsible for at least one of the most spectacular radio relics \citep[e.g.][]{2012MNRAS.425L..76B}. Finally, our model does not fully reproduce the largest linear sizes as functions of the radio power. This may be caused by a too simplistic modelling of the shock dynamics and geometry.

It turns out that the flux distribution as well as the power-mass relation can help constrain several key parameters such as the value of the magnetic field of the relic or its life time. This work underlines the need for a sample of hydrodynamical simulations of binary cluster mergers to make predictions of the shock dynamics across a wide range of masses, all this provided that the equations of magneto-hydrodynamics apply in the dilute plasma in cluster outskirts. In future work, it may be interesting to explore the effect of relics on EoR foreground as in \cite{2019ApJ...879..104L}, as well as  the anomalous radio synchrotron background measured by Arcade2 \citep[e.g.][]{2016JCAP...10..004F}, as already explored for radio halos.

\section*{Acknowledgements}

We thank Gianfranco Brunetti for useful comments.
MB acknowledges support from the Deutsche Forschungsgemeinschaft under Germany's Excellence Strategy - EXC 2121 "Quantum Universe" - 390833306. This work was in part performed at the Aspen Center for Physics, which is supported by National Science Foundation grant PHY-1607611. FV acknowledges financial support from the ERC Starting Grant "MAGCOW", no. 714196.

\appendix

\section{Effect of minimum momentum of relativistic electrons}
 
The use of a stationary slope for the spectrum of the relativistic electrons as well as the somewhat arbitrary choice of the minimum momentum are limitations of our computational approach. However, in this setup it is not straightforward to include a computation of the evolution of electron spectrum as suggested. In our model the electrons get injected continuously for the life-time of the relic, $t_{\rm dur}$. Moreover, they get injected at the shock wave, whose properties, such as the Mach number, velocity or pre-shock density change as the shock propagates outwards. Hence, the injection is time-dependent and we cannot apply a simple analytical treatment but would have to resort to a solution of a Fokker-Planck diffusion-advection equation.

\begin{figure}
\center{\includegraphics[width=\columnwidth]{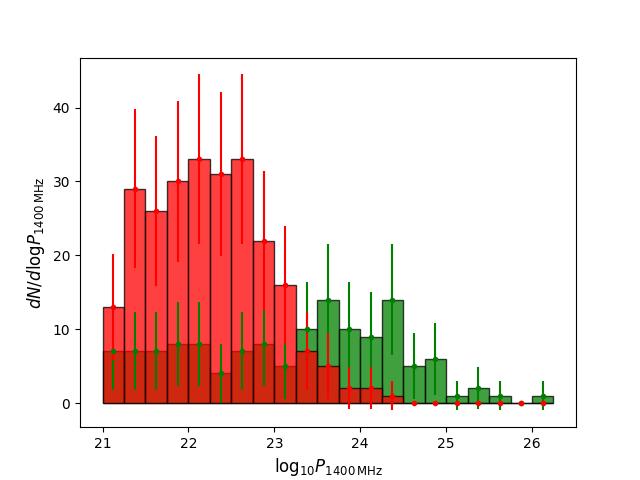}\\
\includegraphics[width=\columnwidth]{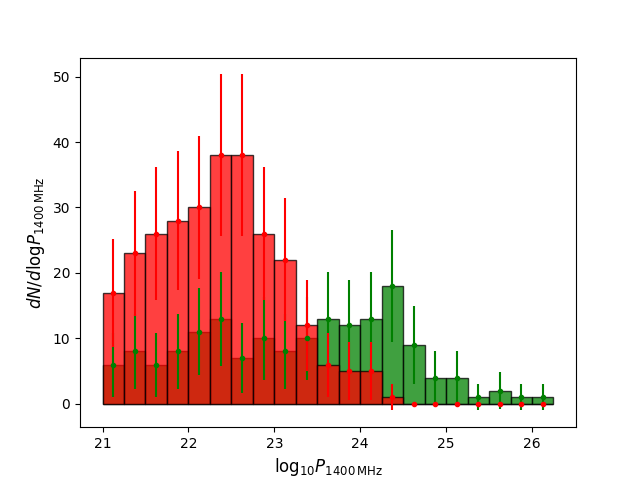}\\
\includegraphics[width=\columnwidth]{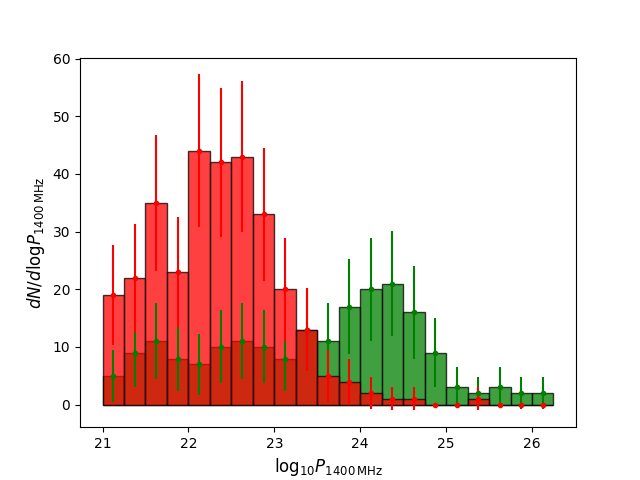}
}
\caption{power distribution at 1.4 GHz for $p_{\rm min}= 1$, 5 and 10 $m_e c$ (from top to bottom).}
\label{fig:pmin}
\end{figure}

In order to assess the dependence on the assumptions of the minimum momentum, we have computed the radio powers at 1400 MHz of runs with minimum momenta, $p_{\rm min}= 1$, 5 and 10 $m_e c$, as these span the values usually used in the literature. While particle-in-cell simulations yield equivalent minimum momenta for relativistic protons, it is true that for electrons there are no real constraints, as part of the injection problem for electrons. The results are shown in Fig.~\ref{fig:pmin}. For higher momenta, the powers of relics increase by a small factor since Eq.~\ref{eq:injection2} shows that $C \propto (s-2)\gamma_{\rm min}^{s-2}$. However, the changes in total numbers are at the level of a few percent.

\bibliographystyle{mnras}
\bibliography{relic} 

\bsp	
\label{lastpage}
\end{document}